\documentclass[11pt]{article}
\usepackage{amssymb,amsmath,amsfonts}
\usepackage{graphicx}
\usepackage{graphics}
\usepackage{eepic,epsfig}

\@addtoreset{equation}{section}

\makeatother

\begin{document}

\title{Current density induced by a cosmic string in de Sitter spacetime in the presence of two flat boundaries}
\author{W. Oliveira dos Santos$^{1}$\thanks{E-mail: wagner.physics@gmail.com}, H. F. Santana Mota$^{1}$\thanks{E-mail: hmota@fisica.ufpb.br} and E. R. Bezerra de Mello$^{1}$\thanks{E-mail: emello@fisica.ufpb.br} \\
%EndAName
\textit{$^{2}$Departamento de F\'{\i}sica, Universidade Federal da Para\'{\i}%
ba}\\
\textit{58.059-970, Caixa Postal 5.008, Jo\~{a}o Pessoa, PB, Brazil}\vspace{%
0.3cm}}
\maketitle

\begin{abstract}
In this paper, we investigate the vacuum  bosonic current density induced by a carrying-magnetic-flux cosmic string in a $(D+1)$-de Sitter spacetime considering the presence of two flat boundaries perpendicular to it. In this setup, the Robin boundary conditions are imposed on the scalar charged quantum field on the boundaries. The particular cases of Dirichlet and Neumann boundary conditions are studied separately. Due to the coupling of the quantum scalar field with the classical gauge field, corresponding to a magnetic flux running along the string's core, a nonzero vacuum expectation value for the current density operator along the azimuthal direction is induced. The two boundaries divide the space in three regions with different properties of the vacuum states. In this way, our main objective is to calculate the induced currents in these three regions. In order to develop this analysis we calculate, for both regions, the positive frequency Wightman functions. Because the vacuum bosonic current in dS space has been investigated before, in this paper we consider only the contributions induced by the boundaries. We show that for each  region the azimuthal current densities are odd functions of the magnetic flux along the string. To probe the correctness of our results, we take the particular cases and analyze some asymptotic limits of the parameters of the model. Also some graphs are presented exhibiting the behavior of the current with relevant physical parameter of the system.
\end{abstract}


\bigskip

\section{Introduction}\label{Int}
De Sitter (dS) space is solution of the Einstein equation in the presence of positive cosmological constant. Although being a curved spacetime it enjoys the same degree of symmetry as the Minkowski one \cite{BD}, so several physical problem can have exact solutions on this background; in addition, the relevance of these theoretical analysis has received great attention due to the appearance of the inflationary cosmology scenario \cite{Linde}. In many inflationary models, an approximate de Sitter (dS) spacetime is used to address relevant problems in standard cosmology. During an inflationary epoch, quantum fluctuations in the inflaton field generate inhomogeneities that can influence the transition to the true vacuum. These fluctuations play a crucial role in the formation of cosmic structures originating from inflation. Specifically the problem of particle creation in the inflationary phase of the Universe, was analyzed in \cite{Mottola} considering de Sitter space. There it is was calculated the energy momentum of the created particles during the inflation, by computing the difference  between the $in$- and $out$-vacuum states

Cosmic strings are linear gravitational topological defects which may have been formed in the early Universe  as  consequence of phase transitions in the context of the standard gauge field theory of elementary particle physics \cite{VS,hindmarsh,Hyde:2013fia}. Although the observations of anisotropies in the Cosmic Microwave Background Radiation by COBE, WMAP and more recently by the Planck Satellite have ruled out cosmic strings as the primary source for primordial density perturbations, they give rise to a number of interesting physical effects such as the emission of gravitational waves and the generation of high-energy cosmic rays (see, for instance, \cite{Damo00}-\cite{Bere}).

The geometry of the spacetime produced by an idealized cosmic string,  i.e., infinitely long and straight,  is locally flat, but topologically conical. It presents a planar angle deficit on the two-surface orthogonal to the string. This object was first introduced in the literature as solution of the Einstein equation in the presence of a Dirac-delta type distribution of energy and axial stress  along a straight infinity line. However this spacetime can also  be obtained in the context of Classical Field Theory, by coupling the energy-momentum tensor associated with a vortex field configuration proposed by Nielsen and Olesen in \cite{Nielsen197345}, with the Einstein's equation, as investigated in \cite{PhysRevD.32.1323} and \cite{Linet1987240}. In both publications, the authors have shown that a planar angle deficit arises on the two-surface perpendicular to a string, as well as a magnetic flux running through its core. The conical geometry in the spacetime produced by a cosmic string has been considered in different lines of research, since the $80's$ of the last century. Gott III in \cite{Gott}, proposed that cosmic strings can produce double images serving as gravitational lens. Also Linet \cite{Linet} has showed that a test charged particle place at rest in the region outside the cosmic string becomes subjected  to repulsive electrostatic self-interaction. In addition, Smith in \cite{Smith}, also has found a similar phenomenon considering gravitational  effect in the Newtonian limit. He sowed that a test massive particle place at rest in the neighborhood of a cosmic string becomes subjected to an  attractive self-interaction. The reason for the last two phenomena resides in the fact that the conical topology produced by a cosmic string distorts the particles fields.

The analysis of the combined effects of the curvature of the dS background and the conical topology produced by the cosmic string  in the vacuum expectation value (VEV) of the induced azimuthal current, $\langle j^\phi\rangle$, associated with a charged scalar field was presented in  \cite{Mohammadi:2020}.
Another type of vacuum polarization arises when boundaries are considered in the system. The imposition of boundary conditions on quantum fields changes the vacuum fluctuations, and result in additional shifts in the VEV of physical
quantities, such as the energy-momentum tensor. In this sense, the investigation of the VEV of the energy-momentum tensor  and the field squared, associated with a charged massive scalar quantum field in the dS background considering the presence of a cosmic string and just one flat plate perpendicular to it, has been developed in \cite{dosSantos:2023}.

In \cite{Elizalde:2010}, the authors  have calculated the VEV of the energy-momentum tensor and the field squared, associated with a massive scalar quantum field propagating in dS spacetime considering the presence of two parallel flat plates. The authors imposed that on the plates, the scalar field obeys Robin boundary condition. Considering this approach they calculated the contributions to these observables, energy-momentum tensor and field squared, induced by the presence of the  plates in the region between them. With the objective to extend these analyses, we decided to consider in this present work the presence of a  carrying-magnetic-flux cosmic string in dS spacetime perpendicular to the two flat plates, and calculate the induced vacuum  current associated with a quantum massive charged scalar field propagating in this manifold. Because the analysis of vacuum bosonic current induced  by a cosmic string in the dS spacetime in absence of flat plates has been developed previously, our focus here is to investigate the contributions induced by the plates. In order to develop these analyses we calculate the Wightman function in this manifold, considering that the bosonic modes are prepared in Bunch-Davies vacuum. Following a procedure similar to \cite{Elizalde:2010}, we decompose this Wightman function in three distinct contributions. One corresponding to the function induced by the cosmic string in dS in absence of plates, plus other two terms induced by the presence of  one plate and two plates, separately. As explained previously, our focus here is to consider the contributions induced by the plates. In this sense we analyze in detail, considering some limiting situations, the only non-vanishing azimuthal components of the boundary induced currents in the three different regions of the space: considering first the Wightman function induced by each plate separately, we obtain the induced current for the corresponding regions outside the plates, and considering the Wightman function induced by the two plate, we calculate the current in the region between the plates. These currents correspond to a Casimir-like effect, i.e., they are induced due to the boundary condition imposition on the quantum fields on the two flat planes, and as we will see, their intensities decay with the distance from the plane, representing a typical Casimir-like effect.

The plan of this work is as follows: In the Section \ref{model}  we present the geometry of the spacetime that we want to consider, the Klein-Gordon equation obeyed by the charged massive quantum field operator, and the boundary condition that the field has to satisfies on the flat plates. The complete set of normalized positive energy solutions of the Klein-Gordon equation in the region between two parallel flat plates considering that the field obeys the Robin boundary condition on them is presented in Section Bpsonic Modes. Having obtained this set of bosonic modes, in the Section \ref{Wight_func} we calculate the corresponding Wightmann function, by adopting the mode sum formula approach. Because the momentum along the direction of cosmic string is discretized, we use the Abel-Plana summation formula to obtain the sum over this quantum number. Doing this procedure, the Wightmann function is expressed as the sum of three contributions, the first one associated with the presence of a cosmic string in the dS space without plates, the second induced by the presence of a single plate, and the third induced by the two plates. The expressions for second function is applied for each plate separately, and the third function only for the region between the plates. In Section \ref{Current} we present formally the complete decomposition of the induced bosonic current, $\langle j^\phi\rangle$. The contribution induced by a single plate is developed in subsection \ref{Azimuthal},  and in the region between plates in \ref{CD-two-plates}.  Also in these subsections, various asymptotic limits of the currents are considered and numerical results are presented. In Section \ref{Conc} we summarize the most relevant results obtained. Throughout the paper, we use natural units $G=\hbar =c=1$.

\section{Background Geometry and Matter Field Content}
\label{model}
The line element describing the geometry produced by a cosmic string in $(1+D)-$de Sitter spacetime is given by the following expression:
\begin{equation}\label{ds1}
	ds^2=dt^2-e^{2t/a}\left(dr^2+r^{2}d\phi^2+dz^{2}+\sum_{i=1}^{D-3}dx_{i}^{2}\right) \ ,
\end{equation}
where $r\ge0$ and $\phi\in[0,2\pi/q]$ define the coordinates on the conical geometry ($q\ge1$ encodes the angle deficit), $(t,z,x_{i})\in(-\infty,\infty)$ and $\alpha$ stands for the length scale of dS spacetime and it is related with the cosmological constant and the curvature scalar, $R$, by the following relations:
\begin{equation}
	\Lambda=\frac{D(D-1)}{2\alpha^2}, \quad R=\frac{D(D+1)}{\alpha^2} \ .
\end{equation}

For convenience of the discussion that follows below the line element \eqref{ds1}, written in synchronous time coordinate, can be expressed in a conformal form by introduction of the conformal time coordinate, $\tau$, defined as $\tau=-\alpha e^{-t/\alpha}$ with $\tau\in(-\infty,0]$. By doing so, we get
\begin{equation}\label{ds2}
	ds^2=\left(\frac{\alpha}{\tau}\right)^2%
	\left(d\tau^2-dr^2-r^{2}d\phi^2-dz^{2}-\sum_{i=1}^{D-3}dx_{i}^{2}\right) \ .
\end{equation}
Note that the line element inside brackets describes an idealized cosmic string in Minkowski spacetime.

In this paper, we want to analyze the vacuum effects due to a propagating charged scalar field in the dS spacetime with a magnetic-carrying-flux cosmic string and in the presence of two flat boundaries. For this purpose we consider the following Klein-Gordon field equation:
\begin{equation}\label{KG}
	(g^{\mu\nu}D_{\mu}D_{\nu}+m^2+\xi R)\varphi(x)=0 \ ,
\end{equation}
where $D_{\mu}=\partial_{\mu}+ieA_{\mu}$ and $m$ is the mass of the scalar field. In addition, in the expression above, we have considered the nonminimal coupling between the background curvature and the scalar field through the term $\xi R$, where $\xi$ is the curvature coupling constant and $R$ denotes the Ricci scalar. The magnetic flux along the string axis is introduced through the vector potential $A_{\mu}=A_{\phi}\delta_{\mu}^{\phi}$, where $A_{\phi}=-q\Phi/2\pi$ is constant and $\Phi$ represents the magnetic flux along the string.

In order to consider the two flat boundaries, we impose that solutions of the Klein-Gordon equation \eqref{KG} satisfy the Robin boundary conditions given by
\begin{equation}\label{RBC}
	(1+\beta_{j}n^{l}\nabla_{l})\varphi(x)=0 \ , \quad z=a_{j}, \ j=1,2,
\end{equation}
where $\beta_{j}$ are constant coefficients (in particular, for $\beta_{j}=0$ and $\beta_{j}=\infty$, the Robin boundary conditions are reduced to the Dirichlet and Neumann boundary conditions, respectively), and $n^{l}$ represents the normal vectors to the boundaries. In the region between the two plates one has $n^{l}=(-1)^{j-1}\delta_{z}^{l}$. According to the notation above, the two flat boundaries are located at $z=a_{1}$ and $z=a_{2}$, with $a_{1}<a_{2}$. Moreover, note that in our setup problem the cosmic string is perpendicular to the two boundaries, since it is located along the $z$-axis.\\
\noindent\\
{\it Bosonic Modes}\\

In this subsection, our aim is to determine the complete set of normalized solutions for the Klein-Gordon equation \eqref{KG}. 

In the spacetime geometry given by \eqref{ds2} and with the gauge field $A_{\mu}=A_{\phi}\delta_{\mu}^{\phi}$, the Klein-Gordon equation simplifies to

	\begin{eqnarray}\label{EQM}
		&&\Bigg[\frac{\partial^2}{\partial\tau^2}%
		+ \frac{(1-D)}{\tau}\frac{\partial}{\partial\tau}%
		+\frac{D(D+1)\xi+(m\alpha)^2}{\tau^2}%
		-\frac{\partial^2}{\partial r^2}%
		-\frac{1}{r}\frac{\partial}{\partial r}%
		-\frac{1}{r^2}\left(\frac{\partial}{\partial\phi}+ieA_{\phi}\right)^{2}%
		\nonumber\\
		&-&\frac{\partial^2}{\partial z^2}%
		-\sum_{i=1}^{D-4}\frac{\partial^2}{\partial(x^{i})^2}\Bigg]\varphi(x)=0 \ .
	\end{eqnarray}

The equation is completely separable and in accordance to the symmetries present in the geometry under consideration; so, we propose the following Ansatz:
\begin{equation}\label{Ansatz}
	\varphi(x)=f(\tau)R(r)h(z)e^{iqn\phi+i\vec{k}\cdot \vec{x}_{||}} \ ,
\end{equation}
where $\vec{x}_{||}$ denotes the coordinates along the $(D-3)$ extra dimensions, with $\vec{k}$ representing the corresponding momenta. The function $h(z)$ will be determined by the Robin boundary conditions that the scalar field satisfies on both flat boundaries placed at $z=a_1$ and $z=a_2$.

Taking the Ansatz proposed above into \eqref{EQM} and admitting that
\begin{equation}\label{DE-z}
	\frac{\partial^2 h(z)}{\partial z^2}=-k_{z}^{2}h(z) \ ,
\end{equation}
we obtain the following differential equations for the functions $f(\tau)$ and $R(r)$:
\begin{equation}\label{DE-tau}
	\Bigg[	\frac{\partial^2}{\partial\tau^2}+\frac{(1-D)}{\tau}\frac{\partial}{\partial\tau}+\frac{D(D+1)\xi+(m\alpha)^2}{\tau^2}+\lambda^{2}\Bigg]f(\tau)=0 \ ,
\end{equation}
and
\begin{equation}\label{DE-r}
	\Bigg[\frac{\partial^2}{\partial r^2}+\frac{1}{r}\frac{\partial}{\partial r}-\frac{q^{2}(n+\alpha)^2}{r^2}+p^{2}\Bigg]R(r)=0 \ ,
\end{equation}
with
\begin{equation}\label{DR}
	\lambda=\sqrt{p^2 + k_{z}^{2} + \vec{k}^2 }
\end{equation}
and the notation
\begin{equation}
	\alpha=\frac{eA_{\phi}}{q}=-\frac{\Phi}{\Phi_{0}} \ ,
	\label{alpha}
\end{equation}
being $\Phi_{0}=2\pi/e$ the quantum flux.

The solution for \eqref{DE-r} that is regular at $r=0$ is given by
\begin{equation}\label{sol-r}
	R(r)=J_{q|n + \alpha|}(pr),
\end{equation}
where $J_{\mu}(x)$ denotes the Bessel function of first kind \cite{Grad}. The solution for the time-dependent equation is expressed by the linear combination of Hankel functions:
\begin{equation}\label{sol-tau}
	f(\tau)=\eta^{D/2}(c_1 H_{\nu}^{(1)}(\lambda \eta) + c_2 H_{\nu}^{(2)}(\lambda \eta)) \ ,
\end{equation}
with the order given by
\begin{equation}\label{order-EQ-tau}
	\nu=\sqrt{D^2/4 - m^2a^2 - \xi D(D+1)} \ .
\end{equation}

Additionally, in \eqref{sol-tau} we have defined the variable $\eta=|\tau|$, and $H_{\nu}^{(l)}(x)$ denotes the Hankel function \cite{Grad}. Different choices of the coefficients $c_{1,2}$ in \eqref{sol-tau} lead to different choices of the vacuum state. In this paper we consider the Bunch-Davies vacuum, corresponding to the choice $c_{2}=0$.

As to the solution of the $z$-dependent equation, it is constrained in the region $a_{1}<z<a_{2}$ by the Robin boundary conditions \eqref{RBC} on the two flat boundaries. For the plate at $z=a_1$, we have
\begin{equation}\label{sol-z}
	h(z)=\cos[k_{z}(z - a_{1})+\alpha_{1}(k_z)] \ ,
\end{equation}
with the notation
\begin{equation}\label{FBC}
	e^{2i\alpha_{1}(x)}=\frac{i\beta_{1}x-1}{i\beta_{1}x+1} \ .
\end{equation}

From the boundary condition on the second plate $z=a_2$, we get the following equation:
\begin{equation}\label{SBC}
	(1 - b_{1}b_{2}v^2)\sin(v)-(b_{1} + b_{2})v\cos(v)=0, \quad v=k_{z}\tilde{a} \ ,
\end{equation}
with $\tilde{a}=a_{2}-a_{1}$ and $b_{j}=\beta_{j}/\tilde{a}$. We will denote the solutions of \eqref{SBC} by $v=v_{l}$, with $l=1 \ , \ 2, \ 3 ...$ These solutions constrain the eigenvalues $k_{z}$ through the relation $k_{z}=v_{l}/\tilde{a}$.

Finally, combining \eqref{sol-r}, \eqref{sol-tau} and \eqref{sol-z}, we obtain the mode functions that satisfy both the Klein-Gordon equation \eqref{KG} and the Robin boundary conditions \eqref{RBC} on the plates:
\begin{equation}\label{MF}
	\varphi_{\sigma}(x)=C_{\sigma}\eta^{D/2}H_{\nu}^{(1)}(\lambda \eta)J_{q|n + \alpha|}(pr)\cos[k_{z}(z - a_{1})+\alpha_{1}(k_z)]e^{iqn\phi+i\vec{k}\cdot \vec{x}_{||}} \ ,
\end{equation}
where $\sigma=\{\lambda, p, n, k_z, \vec{k}\}$ represents the set of quantum numbers that specify each mode of the field. The coefficient $C_{\sigma}$ is fixed by the orthonormalization condition
\begin{equation}\label{NCF}
	-i\int d^{D-1}x\int_{a_1}^{a_2}dz\sqrt{|g|}g^{00}[\varphi_{\sigma}(x)\partial_{t}\varphi_{\sigma^\prime}^{\ast}(x)-\varphi_{\sigma^\prime}^{\ast}(x)\partial_{t}\varphi_{\sigma}(x)]=\delta_{\sigma,\sigma^{\prime}} \ ,
\end{equation}
where the integral is evaluated over the spatial hypersurface $\tau=$ const, and $\delta_{\sigma,\sigma^{\prime}}$ represents the Kronecker-delta for discrete indices and Dirac-delta function for continuous ones. Applying the normalization condition to the mode functions in \eqref{MF} gives
\begin{equation}\label{NC}
	|C_{\sigma}|^{2}=\frac{(2\pi)^{3-D}\alpha^{1-D}qpe^{i(\nu-\nu^{\ast})\pi/2}}{4\tilde{a}\{1+\cos[v_{l}+2\alpha_{1}(v_{l}/\tilde{a})]\sin(v_{l})/v_{l}\}} \ .
\end{equation}

\section{Wightman Function}
\label{Wight_func}

In this paper, our objective is to examine the vacuum polarization effects arising from the background setup described in the previous section. To achieve this, we will utilize the Wightman function, which is particularly useful for calculating vacuum expectation values of physical observables dependent on bilinear field operators. Specifically, the vacuum properties can be characterized by the positive-frequency Wightman function, $W(x,x^{\prime})=\langle 0|\hat{\varphi}(x)\hat{\varphi}^{\ast}(x^{\prime}) |0\rangle$, where $|0\rangle$ denotes the vacuum state. To evaluate this function, we will use the mode-sum technique, expressing the Wightman function in the form:
\begin{equation}
	W(x,x^{\prime})=\sum_{\sigma}\varphi_{\sigma}(x)\varphi_{\sigma}^{\ast}(x^{\prime}) \ .
	\label{WF}
\end{equation}
where $\sum_{\sigma}$ denotes the summation over both discrete and continuous quantum numbers, with $\sigma=\{\lambda, p, n, k_z, \vec{k}\}$.

Taking \eqref{MF}, along with the coefficient \eqref{NC}, into \eqref{WF}, we obtain
\begin{eqnarray}
	W(x,x^{\prime})&=&\frac{4q(\eta \eta^{\prime})^{D/2}}{(2\pi)^{D-1}\alpha^{D-1}\tilde{a}} \sum_{n=-\infty}^{\infty}e^{inq\Delta\phi}\int_{0}^{\infty} dp p J_{q|n + \alpha|}(pr)J_{q|n + \alpha|}(pr^{\prime})\nonumber\\%
	&\times&\int_{-\infty}^{\infty} d\vec{k}e^{i\vec{k}\cdot\Delta\vec{x}_{||}}\sum_{l=-\infty}^{\infty}K_{\nu}(e^{-\pi i/2}\eta \lambda_{l})K_{\nu}(e^{\pi i/2}\eta^{\prime} \lambda_{l})\nonumber\\%
	&\times&\frac{\cos[v_{l}(z - a_{1})/\tilde{a}+\alpha_{1}(v_{l}/\tilde{a})]\cos[v_{l}(z^{\prime} - a_{1})/\tilde{a}+\alpha_{1}(v_{l}/\tilde{a})]}{1+\cos[v_{l}+2\alpha_{1}(v_{l}/\tilde{a})]\sin(v_{l})/v_{l}} \ ,
	\label{WF2}
\end{eqnarray}
where $\Delta \phi=\phi^{\prime}-\phi$ and $\Delta \vec{x}_{||}=\vec{x}_{||}^\prime-\vec{x}_{||}$. Moreover, to obtain the expression above we have introduced the notation $\lambda_{l}=\sqrt{p^{2} + v_{l}^{2}/\tilde{a}^{2} + \vec{k}^{2}}$ and used the identity \cite{Abra},
\begin{equation}
	e^{i(\nu-\nu^{\ast})\pi/2}H_{\nu}^{(1)}(\lambda \eta)\big[H_{\nu}^{(1)}(\lambda \eta^{\prime})\big]^{\ast}=\frac{4}{\pi^2}K_{\nu}(-i\lambda\eta)K_{\nu}(i\lambda\eta^{\prime}) \ .
\end{equation}

To develop the sum over the quantum number $l$, we adopt a variant of the Abel-Plana summation formula \cite{Elizalde:2010},
\begin{eqnarray}
	\sum_{l=1}^{\infty}\frac{\pi v_{l}f(v_{l})}{v_{l}+\sin(v_{l})\cos[v_{l}+2\alpha_{1}(v_{l}/\tilde{a})]}&=&-\frac{\pi}{2}\frac{f(0)}{1-b_1-b_2}+\int_{0}^{\infty}dyf(y)\nonumber\\
	&+&i\int_{0}^{\infty}dy\frac{f(iy)-f(-iy)}{\frac{(b_1y-1)}{b_1+1}\frac{(b_2y-1)}{b_2+1}e^{2y}-1} \ .
	\label{APF}
\end{eqnarray}

For our case,

	\begin{equation}
		f(y)=K_{\nu}(e^{-\pi i/2}\eta \lambda_{l})K_{\nu}(e^{\pi i/2}\eta^{\prime} \lambda_{l})\cos[y(z - a_{1})/\tilde{a}+\alpha_{1}(y/\tilde{a})]\cos[y(z^{\prime} - a_{1})/\tilde{a}+\alpha_{1}(y/\tilde{a})] \ .
	\end{equation}

In accordance with the formula above, the Wightman function can be decomposed as
\begin{equation}
	W(x,x^{\prime})=W_{\rm{1}}(x,x^{\prime})+\Delta W(x,x^{\prime}) \ ,
	\label{Decomp-WF} 
\end{equation}
where the first term corresponds to the contribution to a single plate in $z=a_1$ with a cosmic string perpendicular to it and has been considered in \cite{dosSantos:2023} in the analysis of VEV of the bosonic energy-momentum tensor. It reads,
\begin{eqnarray}
	W_{\rm{1}}(x,x^{\prime})&=&\frac{8q(\eta \eta^{\prime})^{D/2}}{(2\pi)^{D}a^{D-1}} \sum_{n=-\infty}^{\infty}e^{inq\Delta\phi}\int_{0}^{\infty} dp p  J_{q|n + \alpha|}(pr)J_{q|n + \alpha|}(pr^{\prime})\int d\vec{k}e^{i\vec{k}\cdot\Delta\vec{x}_{||}}\nonumber\\%
	&\times&\int_{0}^{\infty}du K_{\nu}(e^{-\pi i/2}\eta \sqrt{u^2+p^2+k^2})K_{\nu}(e^{\pi i/2}\eta^{\prime} \sqrt{u^2+p^2+k^2})\nonumber\\%
	&\times&\cos[u(z - a_{1})+\alpha_{1}(u)]\cos[u(z^{\prime} - a_{1})+\alpha_{1}(u)] \ .
	\label{WF3}
\end{eqnarray}

The second contribution in \eqref{Decomp-WF}, $\Delta W(x,x^{\prime})$ is the interference term and it is given by

	\begin{eqnarray}
		\Delta W(x,x^{\prime})&=&\frac{2q}{(2\pi)^{D-1}a^{D-1}} \sum_{n=-\infty}^{\infty}e^{inq\Delta\phi}\int_{0}^{\infty} dp p  J_{q|n + \alpha|}(pr)J_{q|n + \alpha|}(pr^{\prime})\int d\vec{k}e^{i\vec{k}\cdot\Delta\vec{x}_{||}}\nonumber\\%
		&\times&\int_{\sqrt{p^2+k^2}}^{\infty}\frac{du}{c_1(u)c_2(u)e^{2\tilde{a}u}-1} \cos[u(z - a_{1})+\tilde{\alpha}_{1}(u)]\cos[u(z^{\prime} - a_{1})+\tilde{\alpha}_{1}(u)]\nonumber\\%
		&\times&y^{-D}[\tilde{K}_{\nu}(\eta y)\tilde{I}_{\nu}(\eta^{\prime} y)+\tilde{I}_{\nu}(\eta y)\tilde{K}_{\nu}(\eta^{\prime} y)]\Big|_{y=\sqrt{u^2-p^2-k^2}} \ , 
		\label{WF4}
	\end{eqnarray}
where $\tilde{\alpha}_{1}(u)$ is given by the relation $e^{2\tilde{\alpha}_{1}(u)}=c_1(u)$ and the following notations were introduced:
\begin{equation}
	\tilde{K}_{\nu}=y^{D/2}K_{\nu}(y), \ 	\tilde{I}_{\nu}=y^{D/2}[I_{\nu}(y)+I_{-\nu}(y)] \ ,
\end{equation}
and
\begin{equation}
	c_j(u)=\frac{\beta_ju-1}{\beta_ju+1} \ .
\end{equation}

For further convenience, the contribution induced by a single plate given in \eqref{WF3} can be generalized as
\begin{equation}
	W_{j}(x,x^{\prime})=W_{\rm{dS,cs}}(x,x^{\prime})+W_{j}^{(1)}(x,x^{\prime}) \ ,
	\label{Decomp-WF2}
\end{equation}
where for $j=1$ or $j=2$ it is induced by a single plate at $z=a_1$ or $z=a_2$, respectively. Moreover, the first term in the above expression  contains two contributions: one induced in pure dS spacetime, i.e., in the absence of cosmic string, and the other one induced by it. This contribution reads,

	\begin{eqnarray}
		W_{\rm{dS,cs}}(x,x^{\prime})&=&\frac{4q(\eta \eta^{\prime})^{D/2}}{(2\pi)^{D}a^{D-1}} \sum_{n=-\infty}^{\infty}e^{inq\Delta\phi}\int_{0}^{\infty} dp p  J_{q|n+\alpha|}(pr)J_{q|n+\alpha|}(pr^{\prime})\int d\vec{k}e^{i\vec{k}\cdot\Delta\vec{x}_{||}}\nonumber\\%
		&\times&\int_{0}^{\infty}du K_{\nu}(e^{-\pi i/2}\eta \sqrt{u^2+p^2+k^2})K_{\nu}(e^{\pi i/2}\eta^{\prime} \sqrt{u^2+p^2+k^2})\cos(u\Delta z) \ .
		\label{WF5}
	\end{eqnarray}

In \cite{dosSantos:2023} this function has been explicitly developed.

Our aim in this work is to investigate the current induced by the plates. So, let us consider first the second contribution in \eqref{Decomp-WF2}, which is induced by a single plate at $z=a_j$ and it is given by

	\begin{eqnarray}
		W_{j}^{(1)}(x,x^{\prime})&=&\frac{qa^{1-D}}{2(2\pi)^{D-1}} \sum_{n=-\infty}^{\infty}e^{inq\Delta\phi}\int_{0}^{\infty} dp p  J_{q|n + \alpha|}(pr)J_{q|n + \alpha|}(pr^{\prime})\int d\vec{k}e^{i\vec{k}\cdot\Delta\vec{x}_{||}}\nonumber\\%
		&\times&\int_{\sqrt{p^2+k^2}}^{\infty}du\frac{e^{-u|z+z^{\prime}-2a_j|}}{c_j(u)}y^{-D}[\tilde{K}_{\nu}(\eta y)\tilde{I}_{\nu}(\eta^{\prime} y)+\tilde{I}_{\nu}(\eta y)\tilde{K}_{\nu}(\eta^{\prime} y)]\Big|_{y=\sqrt{u^2-p^2-k^2}} \ .
		\label{WF6}
	\end{eqnarray}

The decomposition in \eqref{Decomp-WF2}, allow us rewrite the Wightman function in the more symmetric form:
\begin{eqnarray}
	W(x,x^{\prime})&=&W_{\rm{dS,cs}}(x,x^{\prime})+\sum_{j=1,2}W_{j}(x,x^{\prime})+\frac{q}{2(2\pi)^{D-1}a^{D-1}} \sum_{n=-\infty}^{\infty}e^{inq\Delta\phi}\nonumber\\%
	&\times&\int_{0}^{\infty} dp p  J_{q|n + \alpha|}(pr)J_{q|n + \alpha|}(pr^{\prime})\int d\vec{k}e^{i\vec{k}\cdot\Delta\vec{x}_{||}}\int_{\sqrt{p^2+k^2}}^{\infty}\frac{du}{c_1(u)c_2(u)e^{2\tilde{a}u}-1}\nonumber\\%
	&\times& \left[2\cosh(u\Delta z)+\sum_{j=1,2}e^{-u|z+z^{\prime}-2a_j|}/c_j(u)\right]\nonumber\\%
	&\times&y^{-D}[\tilde{K}_{\nu}(\eta y)\tilde{I}_{\nu}(\eta^{\prime} y)+\tilde{I}_{\nu}(\eta y)\tilde{K}_{\nu}(\eta^{\prime} y)]\Big|_{y=\sqrt{u^2-p^2-k^2}} \ .
	\label{WF7}
\end{eqnarray}
where the last term is interference part, $\Delta W(x,x^{\prime})$, that is induced by the two plates. For further convenience, we will examine the problem in the particular cases of the well known Dirichlet and Neumann boundary conditions, separately, corresponding to $\beta_j\rightarrow0$ and $\beta_j\rightarrow\infty$, respectively. This allow us to rewrite the interference part as

	\begin{eqnarray}
		\Delta W_{(J)}(x,x^{\prime})&=&\frac{q}{2(2\pi)^{D-1}a^{D-1}} \sum_{n=-\infty}^{\infty}e^{inq\Delta\phi}\int_{0}^{\infty} dp p  J_{q|n + \alpha|}(pr)J_{q|n + \alpha|}(pr^{\prime})\nonumber\\%
		&\times&\int d\vec{k}e^{i\vec{k}\cdot\Delta\vec{x}_{||}}\int_{\sqrt{p^2+k^2}}^{\infty}\frac{du}{e^{2\tilde{a}u}-1}\left[2\cosh(u\Delta z)+\delta_{(J)}\sum_{j=1,2}e^{-u|z+z^{\prime}-2a_j|}\right]\nonumber\\%
		&\times&y^{-D}[\tilde{K}_{\nu}(\eta y)\tilde{I}_{\nu}(\eta^{\prime} y)+\tilde{I}_{\nu}(\eta y)\tilde{K}_{\nu}(\eta^{\prime} y)]\Big|_{y=\sqrt{u^2-p^2-k^2}} \ ,
		\label{WF8}
	\end{eqnarray}
where $J=D$ for Dirichlet BC, $\delta_{(D)}=-1$, and $J=N$ for Neumann BC, $\delta_{(N)}=1$.

Now, introducing a new variable $v=\sqrt{u^2-p^2-k^2}$ and using the identity $(e^{2\tilde{a}u}-1)^{-1}=\sum_{l=1}^{\infty}e^{-2u\tilde{a}l}$, we have

	\begin{eqnarray}
		\Delta W_{(J)}(x,x^{\prime})&=&\frac{q}{2(2\pi)^{D-1}a^{D-1}} \sum_{n=-\infty}^{\infty}e^{inq\Delta\phi}\int_{0}^{\infty} dp p  J_{q|n + \alpha|}(pr)J_{q|n + \alpha|}(pr^{\prime})\int d\vec{k}e^{i\vec{k}\cdot\Delta\vec{x}_{||}}\nonumber\\%
		&\times&\int_{0}^{\infty}\frac{dvv}{\sqrt{v^2+p^2+k^2}}\sum_{l=1}^{\infty}e^{-2\tilde{a}l\sqrt{v^2+p^2+k^2}}\Bigg[2\cosh(\Delta z\sqrt{v^2+p^2+k^2})\nonumber\\%
		&+&\delta_{(J)}\sum_{j=1,2}e^{-|z+z^{\prime}-2a_j|\sqrt{v^2+p^2+k^2}}\Bigg]v^{-D}[\tilde{K}_{\nu}(\eta v)\tilde{I}_{\nu}(\eta^{\prime} v)+\tilde{I}_{\nu}(\eta v)\tilde{K}_{\nu}(\eta^{\prime} v)] \ .
		\label{WF9}
	\end{eqnarray}
We now proceed by using the identity \cite{Grad},
\begin{equation}
	\frac{e^{-ab}}{a}=\frac{2}{\sqrt{\pi}}\int_{0}^{\infty}e^{-a^2s^2-b^2/(4s^2)} \ ,
	\label{id2}
\end{equation}
which allow us to perform the integration over $p$ and $\vec{k}$ variables in \eqref{WF9}, yielding the following result:

%\begin{adjustwidth}{-\extralength}{0cm}
%\centering %% If there is a figure in wide page, please release command \centering
\begin{eqnarray}
	\Delta W_{(J)}(x,x^{\prime})&=&\frac{4q}{(4\pi)^{D/2+1}a^{D-1}}\sum_{l=1}^{\infty}\int_{0}^{\infty}\frac{ds}{s^{D-1}}e^{-(r^2+r^{\prime2}+\Delta\vec{x}_{\parallel}^2)/(4s^2)} \int_{0}^{\infty}dvve^{-s^2v^2}\nonumber\\%
	&\times&\Bigg[\sum_{\epsilon=\pm1}e^{-(2\tilde{a}l+\epsilon\Delta z)^2/(4s^2)}+\delta_{(J)}\sum_{j=1,2}e^{-(2\tilde{a}l+|z+z^{\prime}-2a_j|)^2/(4s^2)}\Bigg]\nonumber\\%
	&\times&v^{-D}[\tilde{K}_{\nu}(\eta v)\tilde{I}_{\nu}(\eta^{\prime} v)+\tilde{I}_{\nu}(\eta v)\tilde{K}_{\nu}(\eta^{\prime} v)]\sum_{n=-\infty}^{\infty}e^{inq\Delta\phi}I_{q|n+\alpha|}\left(\frac{rr^{\prime}}{2s^2}\right) \ .
	\label{WF10}
\end{eqnarray}
%\end{adjustwidth}

Now in order to continue our calculation, we develop the sum over $n$. The parameter $\alpha$ in Eq. \eqref{alpha} can be written in the form
\begin{equation}
	\alpha=n_{0}+\alpha_0, \ \textrm{with}\ |\alpha_0|<\frac{1}{2}  \  ,
	\label{const-2}
\end{equation}
being $n_{0}$ an integer number. This allow us to sum over the quantum number $n$ in \eqref{WF10} by using the formula obtained in \cite{deMello:2014ksa}:
\begin{eqnarray}
	&&\sum_{n=-\infty}^{\infty}e^{iqn\Delta\phi}I_{q|n+\alpha|}(x)=\frac{1}{q}\sum_{k}e^{x\cos(2\pi k/q-\Delta\phi)}e^{i\alpha_0(2\pi k -q\Delta\phi)}\nonumber\\
	&-&\frac{e^{-iqn_{0}\Delta\phi}}{2\pi i}\sum_{j=\pm1}je^{ji\pi q|\alpha_0|}
	\int_{0}^{\infty}dy\frac{\cosh{[qy(1-|\alpha_0|)]}-\cosh{(|\alpha_0| qy)e^{-iq(\Delta\phi+j\pi)}}}{e^{x\cosh{(y)}}\big[\cosh{(qy)}-\cos{(q(\Delta\phi+j\pi))}\big]} \  ,
	\label{summation-formula}
\end{eqnarray}
where $k$ is an integer number varying in the interval
\begin{equation}
	-\frac{q}{2}+\frac{\Delta\phi}{\Phi_{0}}\le k\le \frac{q}{2}+\frac{\Delta\phi}{\Phi_{0}}  \   .
\end{equation}

The substitution of \eqref{summation-formula} in \eqref{WF10}, allow us to write

	\begin{eqnarray}
		\Delta W_{(J)}(x,x^{\prime})&=&\frac{4}{(4\pi)^{D/2+1}a^{D-1}}\sum_{l=1}^{\infty}\int_{0}^{\infty}\frac{ds}{s^{D-1}}\Biggr\{\sum_{k}e^{i\alpha_0(2\pi k -q\Delta\phi)}\nonumber\\%
		&\times&\Bigg[\sum_{\epsilon=\pm1}e^{-[\rho_{kl}+(2\tilde{a}l+\epsilon\Delta z)^2]/(4s^2)}+\delta_{(J)}\sum_{j=1,2}e^{-[\rho_{klj}+(2\tilde{a}l+|z+z^{\prime}-2a_j|)^2]/(4s^2)}\Bigg]\nonumber\\%
		&-&\frac{e^{-iqn_{0}\Delta\phi}}{2\pi i}\sum_{j=\pm1}je^{ji\pi q|\alpha_0|}\int_{0}^{\infty}dy\frac{\cosh{[qy(1-|\alpha_0|)]}-\cosh{(|\alpha_0| qy)e^{-iq(\Delta\phi+j\pi)}}}{\cosh{(qy)}-\cos{(q(\Delta\phi+j\pi))}}\nonumber\\%
		&\times&\Bigg[\sum_{\epsilon=\pm1}e^{-[\rho_{yl}+(2\tilde{a}l+\epsilon\Delta z)^2]/(4s^2)}+\delta_{(J)}\sum_{j=1,2}e^{-[\rho_{ylj}+(2\tilde{a}l+|z+z^{\prime}-2a_j|)^2]/(4s^2)}\Bigg]\Biggl\}
		\nonumber\\%
		&\times&\int_{0}^{\infty}dve^{-s^2v^2}v^{1-D}[\tilde{K}_{\nu}(\eta v)\tilde{I}_{\nu}(\eta^{\prime} v)+\tilde{I}_{\nu}(\eta v)\tilde{K}_{\nu}(\eta^{\prime} v)] \ ,
		\label{WF11}
	\end{eqnarray}
where the following notation was introduced
\begin{eqnarray}
	\rho_{kl}&=&r^2+r^{\prime2}-2rr^{\prime}\cos(2\pi k/q-\Delta \phi)+\Delta\vec{x}_{\parallel}^2 \ , \nonumber\\
	\rho_{y l}&=&r^2+r^{\prime2}+2rr^{\prime}\cosh(y)+\Delta\vec{x}_{\parallel}^2 \ .
\end{eqnarray}
\section{VEV of the Current Density}
\label{Current}
The VEV of the bosonic current density is formally calculated using the Wightman function through the formula
\begin{equation}
	\langle j_{\mu}(x)\rangle=ie\lim_{x^{\prime}\rightarrow x}{(\partial_{\mu}-\partial_{\mu^{\prime}})W(x,x^{\prime})+2ieA_{\mu}W(x,x^{\prime})} \ .
	\label{CD}
\end{equation}

The only non vanishing component in the setup problem under consideration is the one along the azimuthal direction. According to the decomposition made in \eqref{WF7}, we have
\begin{equation}
	\langle j_{\phi}\rangle=\langle j_{\phi}\rangle_{\rm{dS,cs}}+\sum_{j=1,2}\langle j_{\phi}\rangle_{(J)}^{(j)}+\Delta\langle j_{\phi}\rangle_{(J)} \ .
	\label{CD-decomp}
\end{equation}

This component is induced by the presence of the constant potential vector component along the angular direction, $A_{\phi}$, interacting with the scalar field. Although the corresponding field strength vanishes, the nontrivial topology of the string gives rise to Aharonov–Bohm-like effect on the current density along azimuthal direction. As to the other components of the current density, it can be easily checked that they trivially vanish. 

Let us develop and study each term of \eqref{CD-decomp} individually.
\subsection{Azimuthal Current in the Presence of a Single Plate}
\label{Azimuthal}
The first term on the right-hand side is induced by the string, which is obtained by taking \eqref{WF5} into \eqref{CD}:
\begin{eqnarray}
	\langle j_{\phi}\rangle_{\rm{dS,cs}}&=&-\frac{8q^2e\eta^D}{(2\pi)^{D}a^{D-1}} \sum_{n=-\infty}^{\infty}(n+\alpha)\int_{0}^{\infty} dp p  (J_{q|n+\alpha|}(pr))^2\int d\vec{k}\nonumber\\%
	&\times&\int_{0}^{\infty}du K_{\nu}(e^{-\pi i/2}\eta \sqrt{u^2+p^2+k^2})K_{\nu}(e^{\pi i/2}\eta \sqrt{u^2+p^2+k^2}) \ .
	\label{CD-dS-cs}
\end{eqnarray}

This term has been already analyzed in \cite{Mohammadi:2020} for $(1+3)$-dimensions, considering that the string is compactified to a circle. Our aim in this paper, however, is the study of the contributions engendered in the presence of the plates. 

For the contribution induced in the presence of a single plate, we use the representation of the Wightman function \eqref{WF6} for Dirichlet ($c_{j}(u)=-1$) and Neumann ($c_{j}(u)=1$): 

	\begin{eqnarray}
		\langle j_{\phi}\rangle_{(J)}^{(j)}&=&-\frac{2\delta_{(J)}q^2e\eta^{D}}{(2\pi)^{D-1}a^{D-1}} \sum_{n=-\infty}^{\infty}(n+\alpha)\int_{0}^{\infty} dpp(J_{q|n + \alpha|}(pr))^2\int d\vec{k}\nonumber\\%
		&\times&\int_{\sqrt{p^2+k^2}}^{\infty}due^{-2u|z-a_j|} K_{\nu}(\eta y)[I_{\nu}(\eta y)+I_{-\nu}(\eta y)]\Big|_{y=\sqrt{u^2-p^2-k^2}} \ .
		\label{CDpl}
	\end{eqnarray}
Introducing a new variable $v=\sqrt{u^2-p^2-k^2}$ and using the identity given in \eqref{id2}, we get

	\begin{eqnarray}
		\langle j_{\phi}\rangle_{(J)}^{(j)}&=&-\frac{4\delta_{(J)}q^2e\eta^{D}}{\sqrt{\pi}(2\pi)^{D-1}a^{D-1}} \sum_{n=-\infty}^{\infty}(n+\alpha)\int_{0}^{\infty}dse^{-(z-a_j)^2/(2s^2)}\int_{0}^{\infty} dppe^{-s^2p^2}(J_{q|n + \alpha|}(pr))^2\nonumber\\%
		&\times&\int d\vec{k}e^{-s^2k^2}\int_{0}^{\infty}dvve^{-s^2p^2} K_{\nu}(\eta v)[I_{\nu}(\eta v)+I_{-\nu}(\eta v)] \ .
		\label{CDpl2}
	\end{eqnarray}

We can now perform the integrals over $\vec{k}$, $p$ and $v$. In $\vec{k}$ we have $(D-3)$ Gaussian integrals and the integrals over $p$ and $v$ are performed by using the formulas \cite{Grad}: 
\begin{eqnarray}
	\int_{0}^{\infty} dpp e^{-s^2p^2}(J_{\gamma}(p r))^2=\frac{e^{-r^2/(4s^2)}}{2s^2}I_{\gamma}(r^2/(2s^2))
\end{eqnarray}
and
\begin{eqnarray}
	\int_{0}^{\infty}dvve^{-s^2v^2}K_{\nu}(\eta v)[I_{\nu}(\eta v)+I_{-\nu}(\eta v)]=\frac{e^{\eta^2/(2s^2)}}{2s^2}K_{\nu}(\eta^2/(2s^2)) \ .
	\label{IntegralV}
\end{eqnarray}

Substituting the results of these integrations into \eqref{CDpl2} we get,
\begin{eqnarray}
	\langle j_{\phi}\rangle_{(J)}^{(j)}&=&-\frac{\delta_{(J)}q^2e\eta^{D}}{(2\pi)^{D/2+1}a^{D-1}} \int_{0}^{\infty}d\chi \chi^{D/2-1}e^{-\chi[r^2+(z-a_j)^2-\eta^2]/\eta^2}K_{\nu}(\chi)\nonumber\\%
	&\times&\sum_{n=-\infty}^{\infty}(n+\alpha)I_{q|n+\alpha|}(\chi r^2/\eta^2) \ .
	\label{CDpl3}
\end{eqnarray}

The summation over $n$ has been obtained in \cite{Braganca:2014qma} and it is given by the representation
\begin{eqnarray}
	&&\sum_{n=-\infty}^{\infty}(n+\alpha)I_{q|n+\alpha|}(x)=\frac{2x}{q^2}\sideset{}{'}\sum_{k=0}^{[q/2]}\sin(2\pi k/q)\sin(2\pi k\alpha_0)e^{x\cos(2\pi k/q)}\nonumber\\
	&+&\frac{x}{q\pi}
	\int_{0}^{\infty}dy\sinh(y)\frac{e^{-x\cosh(y)}g(q,\alpha_0,y)}{\cosh{(qy)}-\cos{(q\pi)}} \  ,
	\label{summation-formula2}
\end{eqnarray}
with the function
\begin{eqnarray}
	g(q,\alpha_0,y)=\sin(\alpha_0q\pi)\sinh[(1-|	\alpha_0|)qy]-\sinh(q\alpha_0y)\sin[(1-|\alpha_0|)q\pi] \ .
\end{eqnarray}

In addition, the notation $[q/2]$ denotes the integer part of $q/2$, and the prime on the summation symbol over $k$ indicates, that for even values of $q$, the term $k = q/2$ should be taken with the coefficient $1/2$.

The integration over $\chi$ can be performed by using the formula
\begin{equation}
	\int_{0}^{\infty}dxx^{\mu-1}e^{-vx}K_{\nu}(x)=\frac{\sqrt{\pi}2^{\nu}\Gamma(\mu-\nu)\Gamma(\mu+\nu)}{\Gamma(\mu+1/2)(v+1)^{\mu+\nu}}F\Bigg(\mu+\nu,\nu+\frac{1}{2};\mu+1/2;\frac{v-1}{v+1}\Bigg) \ ,
	\label{IntegralP}
\end{equation}
where $F(a,b;c,z)$ represents the hypergeometric function \cite{Grad}.

Thus, taking \eqref{summation-formula2} into \eqref{CDpl3}, we integrate over $\chi$ obtaining the following result
\begin{eqnarray}
	\langle j^{\phi}\rangle_{(J)}^{(j)}&=&-\frac{2\delta_{(J)}e}{(2\pi)^{(D+1)/2}a^{D+1}}\Bigg[\sideset{}{'}\sum_{k=1}^{[q/2]}\sin(2\pi k/q)\sin(2\pi k\alpha_0)F_{\nu}^{D/2+1}(u_{kj})\nonumber\\
	&+&\frac{q}{\pi}
	\int_{0}^{\infty}dy\frac{\sinh(y)g(q,\alpha_0,y)}{\cosh{(qy)}-\cos{(q\pi)}}F_{\nu}^{D/2+1}(u_{yj})\Bigg] \ ,
	\label{CDpl4}
\end{eqnarray}
where we have introduced the function

	\begin{equation}
		F_{\nu}^{D/2+1}(x)=\frac{2^{\nu+1/2}\Gamma(D/2-\nu+1)\Gamma(D/2+\nu+1)}{\Gamma(D/2+3/2)(x+1)^{D/2+\nu+1}}F\Bigg(\frac{D}{2}+\nu+1,\nu+\frac{1}{2};\frac{D+3}{2};\frac{x-1}{x+1}\Bigg)  \  ,
		\label{F-function}
	\end{equation}
variables
\begin{eqnarray}
	\label{u_function}
	u_{kj}&=&2r_p^2s_k^2+2(z_p-a_j/\eta)^2-1 \ , \nonumber\\
	u_{yj}&=&2r_p^2c_y^2+2(z_p-a_j/\eta)^2-1 \ .
\end{eqnarray}
and the notations
\begin{eqnarray}
	\label{sc}
	s_k=\sin(\pi k/q), \ \ c_y=\cosh(y) \ .
\end{eqnarray}

Moreover, in \eqref{CDpl4}, $r_p=r/\eta$ and $z_p=z/\eta$ are the proper distances from the string and the plate,
respectively, in unities of the dS spacetime curvature, $a$. From \eqref{CDpl4}, we can see that $\langle j^{\phi}\rangle_{(J)}^{(j)}$ is an odd function of
$\alpha_0$ with period equal to quantum flux, $\Phi_{0}=2\pi/e$; moreover,
for $1\leq q<2$, the first term on the right-hand side of \eqref{CDpl4} is absent.

Let us now analyze the behavior of this VEV in some limiting cases. In the conformal coupled massless scalar field case, the function $F_{\nu}^{D/2+1}(x)$ takes the form \cite{dosSantos:2023}:
\begin{equation}
	F_{\nu}^{D/2+1}(x)=\frac{\Gamma(D/2+1/2)}{(x+1)^{(D+1)/2}} \ .
	\label{F-conformal}
\end{equation}

Therefore, the azimuthal current density \eqref{CDpl4} in this case reads
\begin{eqnarray}
	\langle j^{\phi}\rangle_{(J)}^{(j)}&=&-\frac{2\delta_{(J)}e\Gamma(D/2+1/2)}{(4\pi)^{(D+1)/2}a^{D+1}} \Bigg[\sideset{}{'}\sum_{k=1}^{[q/2]}\frac{\sin(2\pi k/q)\sin(2\pi k\alpha_0)}{[r_p^2s_k^2+(z_p-a_j/\eta)^2]^{(D+1)/2}}\nonumber\\
	&+&\frac{q}{\pi}
	\int_{0}^{\infty}dy\frac{\sinh(y)g(q,\alpha_0,y)}{[\cosh{(qy)}-\cos{(q\pi)][r_p^2c_y^2+(z_p-a_j/\eta)^2]^{(D+1)/2}}}\Bigg] \ .
	\label{CDplm0}
\end{eqnarray}

We now want to study the asymptotic behavior of the azimuthal current density in the limits near and distant points from the core of the string, located at $r=0$. For points outside of the plate, $z\neq a_j$, the VEV of the current density on the string is finite for $q|\alpha_0|>1$ and
can be obtained directly by putting $r=0$ in \eqref{CDpl4}. On the other hand, for $q|\alpha_0|<1$, the VEV diverges near the string as
\begin{eqnarray}
	\langle j^{\phi}\rangle_{(J)}^{(j)}&\approx&-\frac{2^{q|\alpha_0|-1/2}\delta_{(J)}qe\Gamma(\frac{3}{2}-q|\alpha_0|)}{(2\pi)^{(D+3)/2}a^{D+1}r_p^{2(1-q|\alpha_0|)}}F_{\nu}^{D/2+q|\alpha_0|-1/2}(2(z_p-a_j/\eta)^2-1) \ .
	\label{CDpl-approx}
\end{eqnarray}

For distant points from the string, $r\gg \eta,|z-a_j|$, we use the corresponding asymptotic expression for function $F_{\nu}^{D/2+1}(x)$ \cite{dosSantos:2023}:
\begin{equation}
	F_{\nu}^{D/2+1}(x)\approx\frac{2^{\nu-1/2}}{\sqrt{\pi}}\Gamma(\nu)\Gamma(D/2-\nu+1)x^{-D/2-1+\nu} \ ,
	\label{asymp}
\end{equation}
which taken into \eqref{CDpl4} gives
\begin{eqnarray}
	\langle j^{\phi}\rangle_{(J)}^{(j)}&\approx&-\frac{2^{2\nu-D-1}\delta_{(J)}e\Gamma(\nu)\Gamma(D/2-\nu+1)}{\pi^{D/2+1}a^{D+1}r_p^{D+2-2\nu}}\Bigg[\sideset{}{'}\sum_{k=1}^{[q/2]}\frac{\sin(2\pi k/q)\sin(2\pi k\alpha_0)}{s_k^{D+2-2\nu}}\nonumber\\
	&+&\frac{q}{\pi}
	\int_{0}^{\infty}dy\frac{\sinh(y)g(q,\alpha_0,y)}{c_y^{D+2-2\nu}[\cosh{(qy)}-\cos{(q\pi)]}}\Bigg] \ .
	\label{CDpl-large-r}
\end{eqnarray}

We now turn to the investigation of the behavior of azimuthal current density for points close and far from the plate at $z=a_j$. On the surface of the plate, $z=a_j$, the VEV induced by the plate is finite for points outside the string, $r\neq0$. However, for points on the string's, $r=0$, and $q|\alpha_0|>1$, the VEV diverges as
\begin{eqnarray}
	\langle j^{\phi}\rangle_{(J)}^{(j)}&\approx&-\frac{4\delta_{(J)}e\Gamma(\frac{D+1}{2})}{(4\pi)^{(D+1)/2}a^{D+1}|z_p-a_j/\eta|^{D-1}}\Bigg[\sideset{}{'}\sum_{k=1}^{[q/2]}\sin(2\pi k/q)\sin(2\pi k\alpha_0)\nonumber\\
	&+&\frac{q}{\pi}
	\int_{0}^{\infty}dy\frac{\sinh(y)g(q,\alpha_0,y)}{\cosh{(qy)}-\cos{(q\pi)}}\Bigg] \ ,
	\label{CDpl-near-aj}
\end{eqnarray}
and for $q|\alpha_0|<1$, it diverges as
\begin{eqnarray}
	\langle j^{\phi}\rangle_{(J)}^{(j)}&\approx&-\frac{4\delta_{(J)}qe\Gamma(\frac{D+1}{2})\Gamma(\frac{3}{2}-q|\alpha_0|)}{(2\pi)^{(D+3)/2}a^{D+1}r_p^{2(1-q|\alpha_0|)}|z_p-a_j/\eta|^{D+2q|\alpha_0|-4}} \ .
	\label{CDpl-near-aj2}
\end{eqnarray}

For distant regions from the plate, $|z-a_j|\gg\eta,r$, we use again the formula given in \eqref{asymp}, obtaining
\begin{eqnarray}
	\langle j^{\phi}\rangle_{(J)}^{(j)}&\approx&-\frac{2^{2\nu-D-1}\delta_{(J)}e\Gamma(\nu)\Gamma(D/2-\nu+1)}{\pi^{D/2+1}a^{D+1}|z_p-a_j/\eta|^{D+2-2\nu}}\Bigg[\sideset{}{'}\sum_{k=1}^{[q/2]}\sin(2\pi k/q)\sin(2\pi k\alpha_0)\nonumber\\
	&+&\frac{q}{\pi}
	\int_{0}^{\infty}dy\frac{\sinh(y)g(q,\alpha_0,y)}{[\cosh{(qy)}-\cos{(q\pi)}][1+r_p^2c_y^2/(z_p-a_j/\eta)^2]^{D/2+1-\nu}}\Bigg] \ .
	\label{CDpl-large-z}
\end{eqnarray}

We also consider the Minkowskian limit, $a\rightarrow\infty$, while keeping $t$ fixed. In this limit, the geometry under consideration simplifies to that of a cosmic string in the background of $(D + 1)$-dimensional Minkowski spacetime. For the analysis of this limit, the representation for the VEV in the presence of one plate, given in \eqref{CDpl4}, is not convenient. Therefore, for this end, we return to the representation present in \eqref{CDpl3} with the series over $n$ given by \eqref{summation-formula2}. For the coordinate $\eta$ in the arguments of the modified Bessel function, we have $\eta\approx |t-a|$. In this limit, $\nu\gg1$ and, according to \eqref{order-EQ-tau}, we have $\nu \approx ima$. Using the uniform asymptotic expansion for the Macdonald function of imaginary order as provided in \cite{Balogh1967}, substituting it into \eqref{CDpl3}, and following some intermediate steps, we obtain
\begin{eqnarray}
	\langle j^{\phi}\rangle_{(J)}^{(j),(M)}&=&-\frac{4\delta_{(J)}em^{D+1}}{(2\pi)^{(D+1)/2}}\Bigg[\sideset{}{'}\sum_{k=1}^{[q/2]}\sin(2\pi k/q)\sin(2\pi k\alpha_0)f_{\frac{D+1}{2}}(2m\sqrt{r^2s_k^2+(z-a_j)^2})\nonumber\\
	&+&\frac{q}{\pi}
	\int_{0}^{\infty}dy\frac{\sinh(y)g(q,\alpha_0,y)}{\cosh{(qy)}-\cos{(q\pi)}}f_{\frac{D+1}{2}}(2m\sqrt{r^2c_y^2+(z-a_j)^2})\Bigg] \ ,
	\label{CDpl-Mink}
\end{eqnarray}
where we have introduced the notation
\begin{equation}
	f_{\mu}(x)=\frac{K_{\mu}(x)}{x^{\mu}} \ ,
	\label{f-function}
\end{equation}
being $K_{\mu}(x)$ the Macdonald function.

\begin{figure}[!htb]
	\begin{center}
		\includegraphics[scale=0.4]{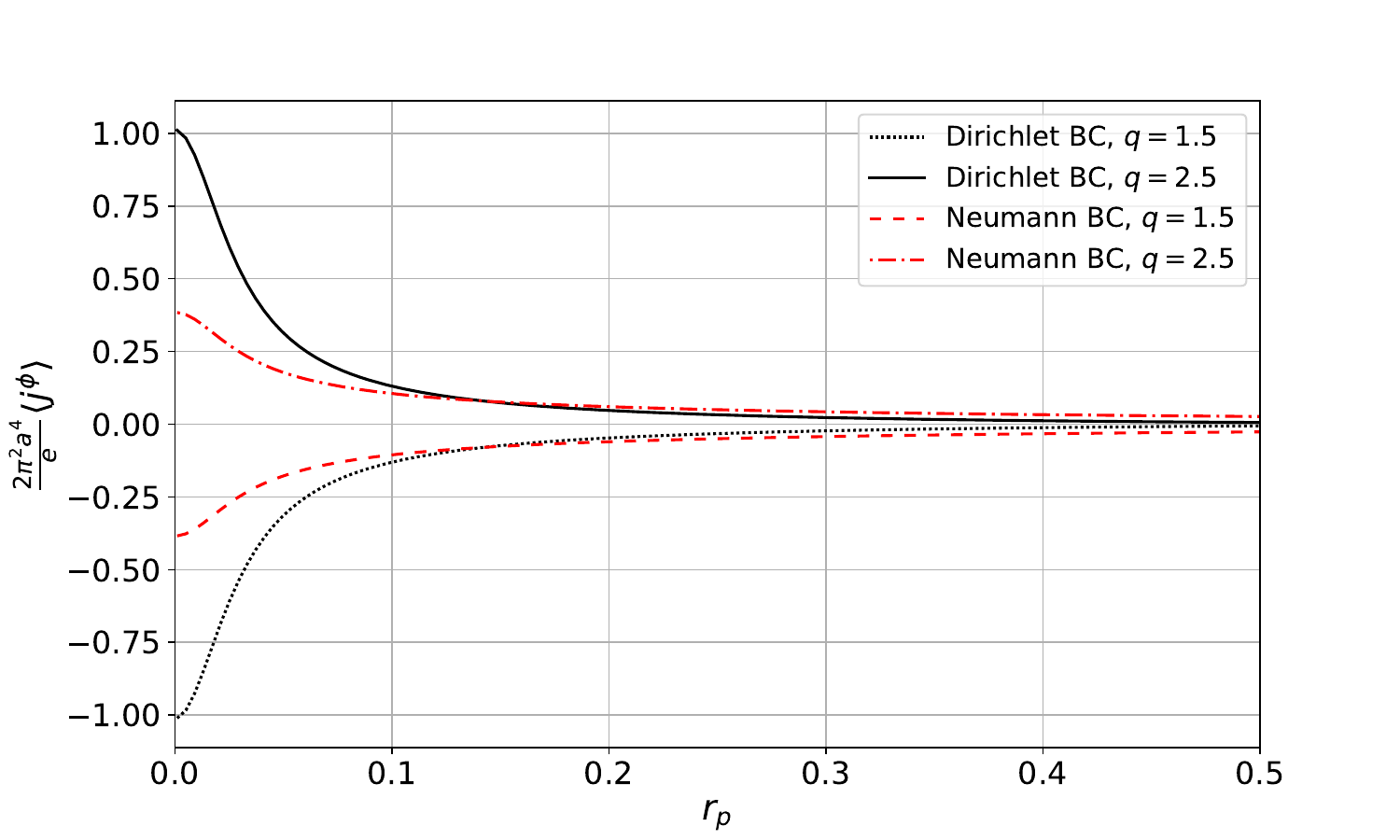}
		%\quad
		\includegraphics[scale=0.4]{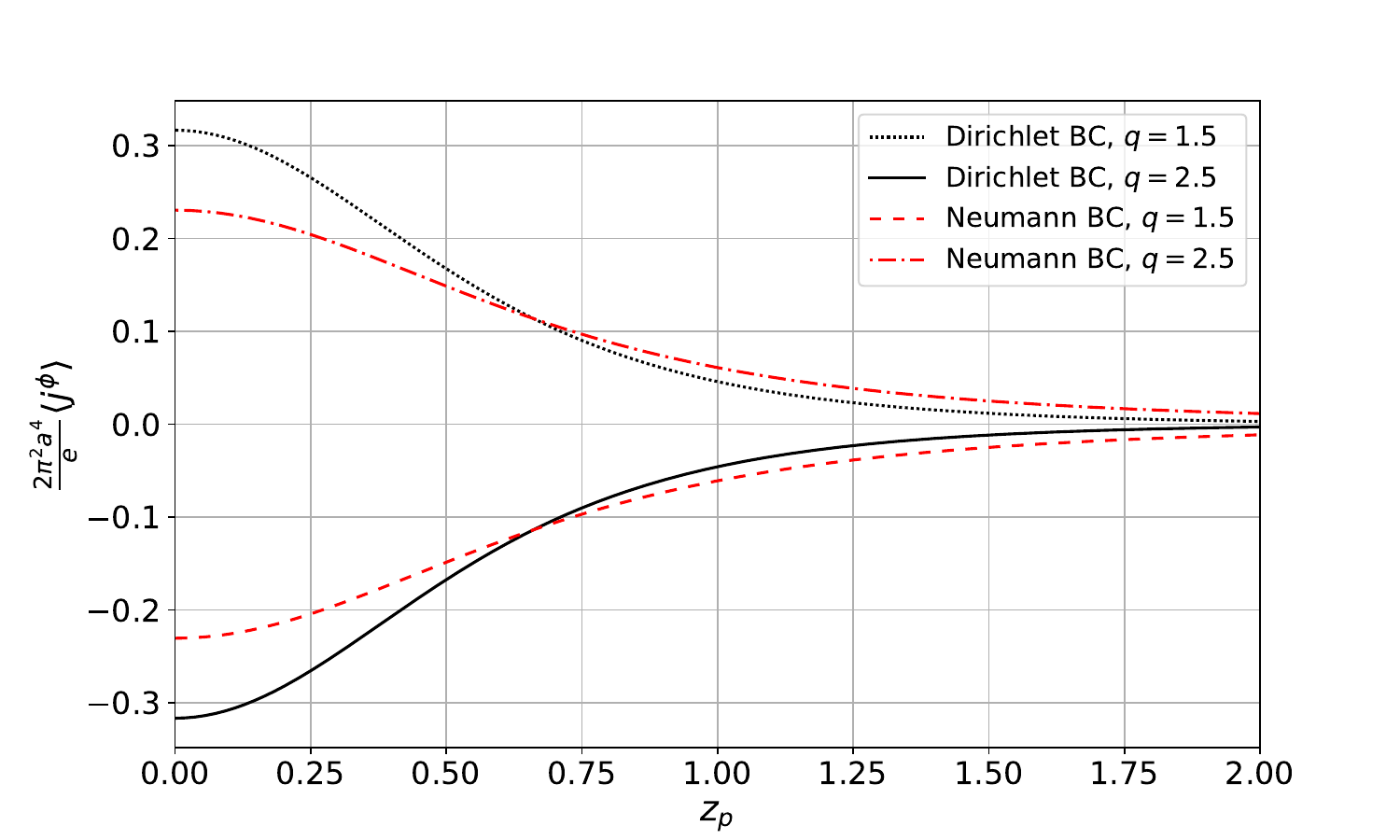}
	\caption{The VEV of the azimuthal current density induced by a single plate, located at $z=a_j$,  is plotted as function of the proper distance from the string, $r_p$, (top panel) and the proper distance from the plate, $z_p$, (bottom panel), in units of $a$. In both plots, we consider Dirichlet and Neumann boundary conditions and various values of $q$. Both graphs are plotted for $D=3$, $\alpha_0=0.25$,  $\xi=0$, $ma=1.5$ and $a_j=0$. Moreover, in the top panel we have fixed $z_p=0$ and in the bottom one, $r_p=1$.}
	\label{fig1}
	\end{center}
\end{figure}

In Fig. \ref{fig1} is exhibited the behavior of the azimuthal current density induced by a single plate at $a_j=0$, as function of the the proper distances from the string, $r_p$, (top panel) and the plate, $z_p$, (bottom panel), in unities of the dS spacetime curvature, $a$. We consider Dirichlet and Neumann boundary conditions for various values of the parameter $q$, which is associated with the deficit angle. From the top panel we can see that the VEV of the azimuthal current density is is finite on the string and rapidly tends to zero as $r_p$ increases. From the bottom panel, we observe that the VEV is finite on the plate location and rapidly goes to zero as $z_p$ goes large, in accordance to our asymptotic analysis. Moreover, note that in both plots the intensities increase with $q$ and are higher for Dirichlet BC, compared with Neumann BC, near the string or the plate.
\subsection{Azimuthal Current  in the Region between the Plates}
\label{CD-two-plates}
Let us analyze now the contribution induced in the region between the plates, $a_1<z<a_2$. To this end, we take \eqref{WF10} into \eqref{CD}, obtaining the expression:
\begin{eqnarray}
	\Delta\langle j_{\phi}\rangle_{(J)}&=&-\frac{16eq^2\eta^2}{(4\pi)^{D/2+1}a^{D-1}}\sum_{l=1}^{\infty}\int_{0}^{\infty}\frac{ds}{s^{D-1}}e^{-r^2/s^2}\nonumber\\%
	&\times&\Bigg[2e^{-(\tilde{a}l)^2/s^2}+\delta_{(J)}\sum_{j=1,2}e^{-(\tilde{a}l+|z-a_j|)^2/s^2}\Bigg]\int_{0}^{\infty}dvve^{-s^2v^2}\nonumber\\%
	&\times&K_{\nu}(\eta v)[I_{\nu}(\eta v)+I_{-\nu}(\eta v)]\sum_{n=-\infty}^{\infty}(n+\alpha)I_{q|n+\alpha|}\left(\frac{r^2}{2s^2}\right) \ .
	\label{CDInt}
\end{eqnarray}

The next step is to integrate over $v$ by using \eqref{IntegralV}:

	\begin{eqnarray}
		\Delta\langle j_{\phi}\rangle_{(J)}&=&-\frac{2q^2e}{(2\pi)^{D/2+1}a^{D-1}} \sum_{l=1}^{\infty}\int_{0}^{\infty}d\chi \chi^{D/2-1}e^{-(r^2/\eta^2-1)\chi}K_{\nu}(\chi)\nonumber\\%
		&\times&\Bigg[2e^{-2\chi(\tilde{a}l/\eta)^2}+\delta_{(J)}\sum_{j=1,2}e^{-2\chi(\tilde{a}l+|z-a_j|)^2/\eta^2}\Bigg]\sum_{n=-\infty}^{\infty}(n+\alpha)I_{q|n+\alpha|}(\chi r^2/\eta^2) \ ,
		\label{CDInt2}
	\end{eqnarray}
where we have introduced the variable $\chi=\eta^2/(2s^2)$.
By using the formula \eqref{summation-formula2} for the sum over $n$, we can integrate over $\chi$ with the help of \eqref{IntegralP}. The result is the following expression:

	\begin{eqnarray}
		\Delta\langle j^{\phi}\rangle_{(J)}&=&-\frac{2e}{(2\pi)^{(D+1)/2}a^{D+1}} \sum_{l=1}^{\infty}\Biggl\{\sideset{}{'}\sum_{k=1}^{[q/2]}\sin(2\pi k/q)\sin(2\pi k\alpha_0)\nonumber\\%
		&\times&\Bigg[2F_{\nu}^{D/2+1}(v_{kl})+\delta_{(J)}\sum_{j=1,2}F_{\nu}^{D/2+1}(v_{kjl})\Bigg]+\frac{q}{\pi}
		\int_{0}^{\infty}dy\frac{\sinh(y)g(q,\alpha_0,y)}{\cosh{(qy)}-\cos{(q\pi)}}\nonumber\\
		&\times&\Bigg[2F_{\nu}^{D/2+1}(v_{yl})+\delta_{(J)}\sum_{j=1,2}F_{\nu}^{D/2+1}(v_{yjl})\Bigg]\Biggr\} \ ,
		\label{CDInt3}
	\end{eqnarray}
where we have introduced the variables
\begin{eqnarray}
	\label{v_variable_1}
	v_{kl}&=&2r_p^2s_k^2+2(\tilde{a}l/\eta)^2-1 \ , \nonumber\\
	v_{yl}&=&2r_p^2c_y^2+2(\tilde{a}l/\eta)^2-1 \ ,
\end{eqnarray}
and
\begin{eqnarray}
	\label{v_variable_2}
	v_{kjl}&=&2r_p^2s_k^2+2(\tilde{a}l/\eta+|z_p-a_j/\eta|)^2-1 \ , \nonumber\\
	v_{yjl}&=&2r_p^2c_y^2+2(\tilde{a}l/\eta+|z_p-a_j/\eta|)^2-1 \ .
\end{eqnarray}

Let us now study the behavior of this VEV in some limiting situations. In the conformal coupled massless scalar field case, the function $F_{\nu}^{D/2+1}(u)$ has the simple form given in \eqref{F-conformal}. Thus,  in this case, the VEV induced in the region between the plates reads:

	\begin{eqnarray}
		\Delta\langle j^{\phi}\rangle_{(J)}&=&-\frac{2e\Gamma(D/2+1/2)}{(4\pi)^{(D+1)/2}a^{D+1}} \sum_{l=1}^{\infty}\Biggl\{\sideset{}{'}\sum_{k=1}^{[q/2]}\sin(2\pi k/q)\sin(2\pi k\alpha_0)\Bigg[\frac{2}{[r_p^2s_k^2+(\tilde{a}l/\eta)^2]^{(D+1)/2}}\nonumber\\%
		&+&\sum_{j=1,2}\frac{\delta_{(J)}}{[r_p^2s_k^2+(\tilde{a}l/\eta+|z_p-a_j/\eta|)^2]^{(D+1)/2}}\Bigg]+\frac{q}{\pi}
		\int_{0}^{\infty}dy\frac{\sinh(y)g(q,\alpha_0,y)}{\cosh{(qy)}-\cos{(q\pi)}}\nonumber\\%
		&\times&\Bigg[\frac{2}{[r_p^2c_y^2+(\tilde{a}l/\eta)^2]^{(D+1)/2}}+\sum_{j=1,2}\frac{\delta_{(J)}}{[r_p^2c_y^2+(\tilde{a}l/\eta+|z_p-a_j/\eta|)^2]^{(D+1)/2}}\Bigg]\Biggr\} \ .
		\label{Conformal-two-plates}
	\end{eqnarray}

We now consider the limit of large values of the distance between the plates, $\tilde{a}\gg r,|z-a_j|$. In this case, we can not neglect the term $2r_p^2c_y^2$ in the arguments of the functions $F_{\nu}^{D/2+1}(u)$, since it is essential for the convergence of the integral over $y$. Therefore, in this limit, we get

	\begin{eqnarray}
		\Delta\langle j^{\phi}\rangle_{(J)}&\approx&\frac{2^{2\nu-D}e\Gamma(\nu)\Gamma(D/2-\nu+1)}{\pi^{D/2+1}a^{D+1}(\tilde{a}/\eta)^{D+2-2\nu}} \sum_{l=1}^{\infty}\Biggl\{\sideset{}{'}\sum_{k=1}^{[q/2]}\sin(2\pi k/q)\sin(2\pi k\alpha_0)\nonumber\\&\times &\left[\frac{2}{l^{D+2-2\nu}}+\sum_{j=1,2}\frac{\delta_{(J)}}{(l+|z-a_j|/\tilde{a})^{D+2-2\nu}}\right]+\frac{q}{\pi}
		\int_{0}^{\infty}dy\frac{\sinh(y)g(q,\alpha_0,y)}{[\cosh{(qy)}-\cos{(q\pi)]}}\nonumber\\&\times &\left[\frac{2}{[(rc_y/\tilde{a})^2+l^2]^{D/2+1-\nu}}+\sum_{j=1,2}\frac{\delta_{(J)}}{[(rc_y/\tilde{a})^2+(l+|z-a_j|/\tilde{a})^2]^{D/2+1-\nu}}\right]\Biggr\} \ .
		\label{Limit_a_large}
	\end{eqnarray}

This result show us that the VEV of the azimuthal current density decays as the separation between the plates increases.

For distant points from the string and fixed distances from the plates, $r\gg \eta,|z-a_j|$, we have $v_{kjl}\approx v_{kl}$ and $v_{yjl}\approx v_{yl}$, according to \eqref{v_variable_1} and \eqref{v_variable_2}. Therefore, using the corresponding asymptotic expression for the function $F_{\nu}^{D/2+1}(u)$ given in \eqref{asymp}, we obtain the following result:

	\begin{eqnarray}
		\Delta\langle j^{\phi}\rangle_{(J)}&\approx&-\frac{2^{2\nu+1-D}(1+\delta_{(J)})e\Gamma(\nu)\Gamma(D/2-\nu+1)}{\pi^{D/2+1}a^{D+1}r_p^{D+2-2\nu}} \sum_{l=1}^{\infty}\Biggl\{\sideset{}{'}\sum_{k=1}^{[q/2]}\frac{\sin(2\pi k/q)\sin(2\pi k\alpha_0)}{[s_k^2+(l\tilde{a}/r)^2]^{D/2+1-\nu}}\nonumber\\
		&+&\frac{q}{\pi}
		\int_{0}^{\infty}dy\frac{\sinh(y)g(q,\alpha_0,y)}{[\cosh{(qy)}-\cos{(q\pi)][c_y^2+(\tilde{a}l/r)^2]^{D/2+1-\nu}}}\Biggr\} \ .
		\label{Limit-large-r}
	\end{eqnarray}

Finally, we consider the Minkowskian limit, $a\rightarrow\infty$, with a fixed value of $t$. Following the same procedure adopted for the contribution induced by a single plate, the VEV of the azimuthal current density induced in the region between the plates reads:

	\begin{eqnarray}
		\Delta\langle j^{\phi}\rangle_{(J)}^{(M)}&=&-\frac{4em^{D+1}}{(2\pi)^{(D+1)/2}a^{D+1}} \sum_{l=1}^{\infty}\Biggl\{\sideset{}{'}\sum_{k=1}^{[q/2]}\sin(2\pi k/q)\sin(2\pi k\alpha_0)\nonumber\\%
		&\times&\Bigg[2f_{\frac{D+1}{2}}(2m\sqrt{r^2s_k^2+(z-a_j)^2})\nonumber\\%
		&+&\delta_{(J)}\sum_{j=1,2}f_{\frac{D+1}{2}}(2m\sqrt{r_p^2s_k^2+(\tilde{a}l/\eta+|z_p-a_j/\eta|)^2})\Bigg]\nonumber\\%
		&+&\frac{q}{\pi}
		\int_{0}^{\infty}dy\frac{\sinh(y)g(q,\alpha_0,y)}{\cosh{(qy)}-\cos{(q\pi)}}\Bigg[2f_{\frac{D+1}{2}}(2m\sqrt{r^2c_y^2+(z-a_j)^2})\nonumber\\%
		&+&\delta_{(J)}\sum_{j=1,2}f_{\frac{D+1}{2}}(2m\sqrt{r_p^2c_y^2+(\tilde{a}l/\eta+|z_p-a_j/\eta|)^2})\Bigg]\Biggr\} \ ,
		\label{Two-plates-Mink}
	\end{eqnarray} 
with the function $f_{\mu}(x)$ defined in \eqref{f-function}.

\begin{figure}[!htb]
	\begin{center}
		\includegraphics[scale=0.3]{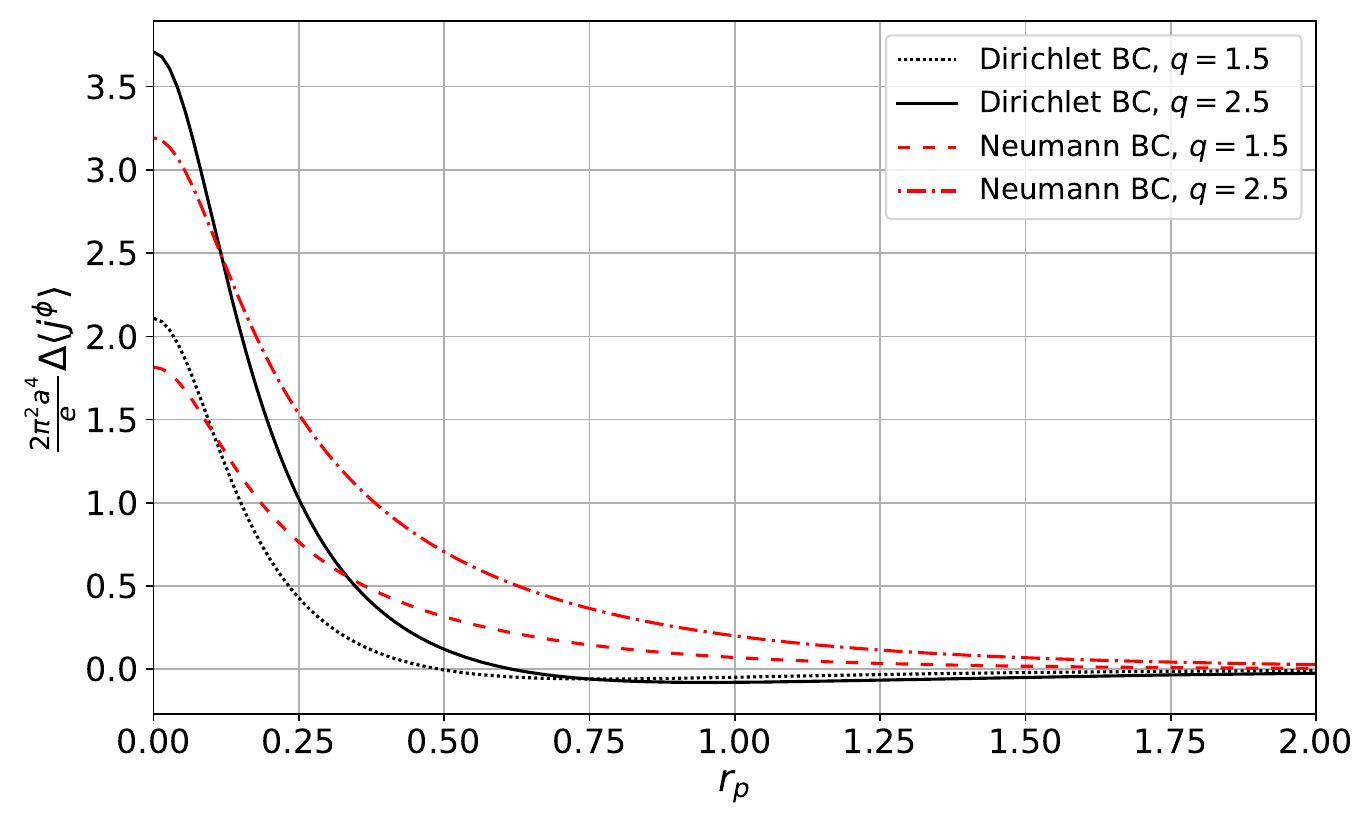}
		\quad
		\includegraphics[scale=0.3]{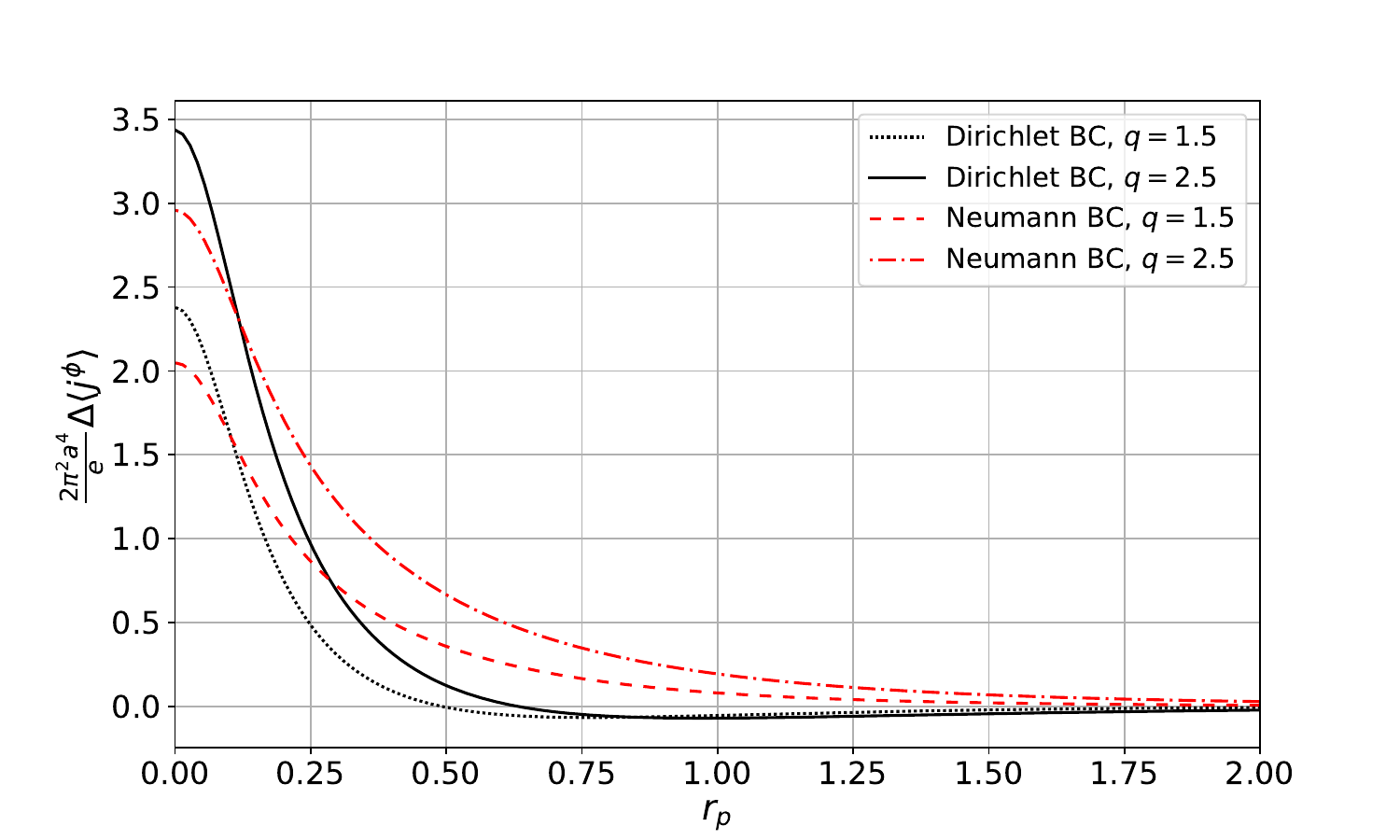}
	\caption{The VEV of the azimuthal current density induced between the plates is plotted as function of the proper distance from the string, $r_p$. In the top panel we consider $z_p=0.2$ and in the bottom panel, $z_p=0.5$. In both plots we consider Dirichlet and Neumann boundary conditions and different values of $q$. Both graphs are plotted for $D=3$, $\alpha_0=0.25$,  $\xi=0$, $ma=1.5$. The positions of the plates are in both plots at $a_1=0$ and $a_2=1$.}
	\label{fig2}
	\end{center}
\end{figure}

In Fig. \ref{fig2} is displayed the dependence of the VEV of the azimuthal current density in the region between the plates as function of the proper distance from the string, $r_p$, considering $z_p=0.2$ (top panel) and $z_p=0.5$ (bottom panel). This is presented in unities of the dS spacetime curvature, $a$. 
In both plots, we consider Dirichlet and Neumann boundary conditions for various values of the parameter associated with the deficit angle, $q$. We observe that the current density in the region between the plates is finite on the string and rapidly goes to zero as the proper distance from the string, $r_p$, increases.

\begin{figure}[!htb]
	\begin{center}
		\includegraphics[scale=0.3]{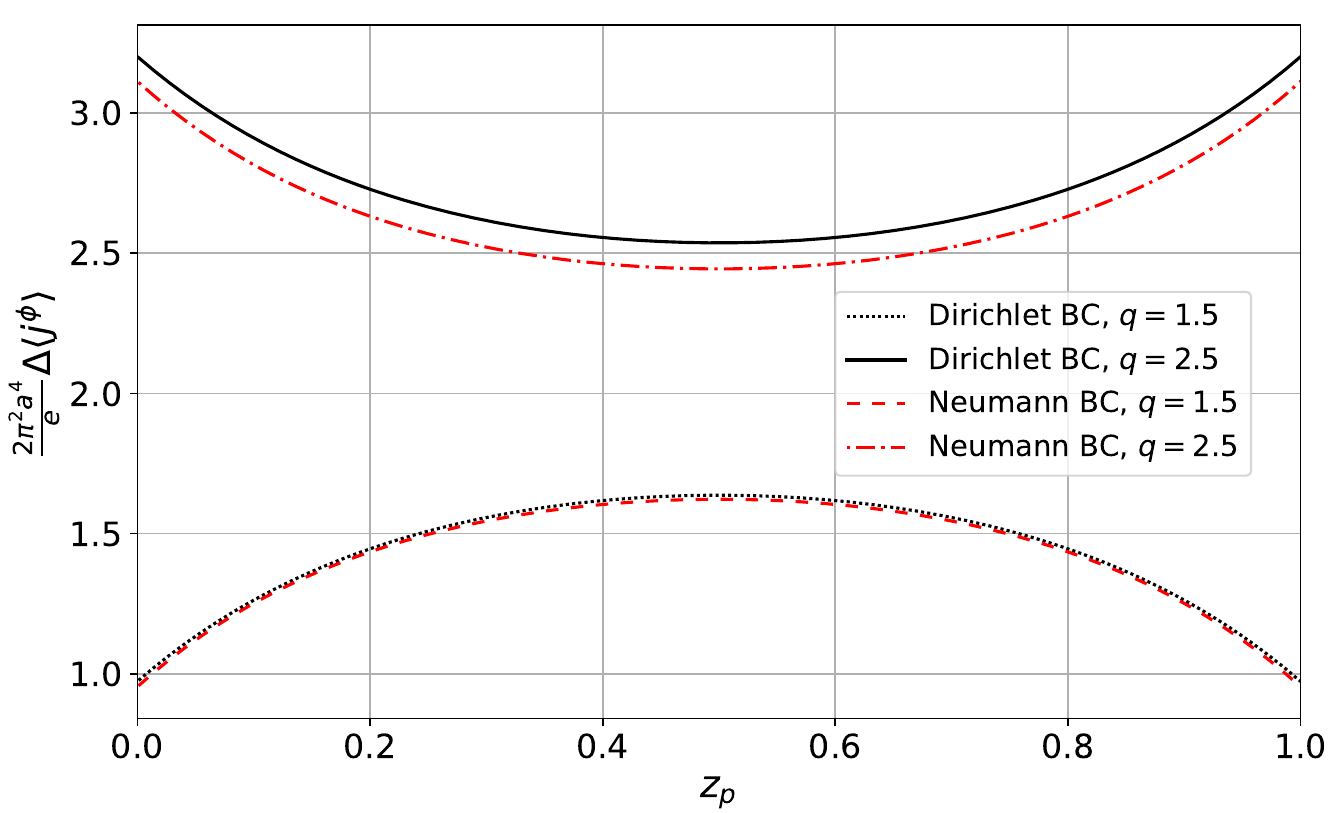}
		\quad
		\includegraphics[scale=0.3]{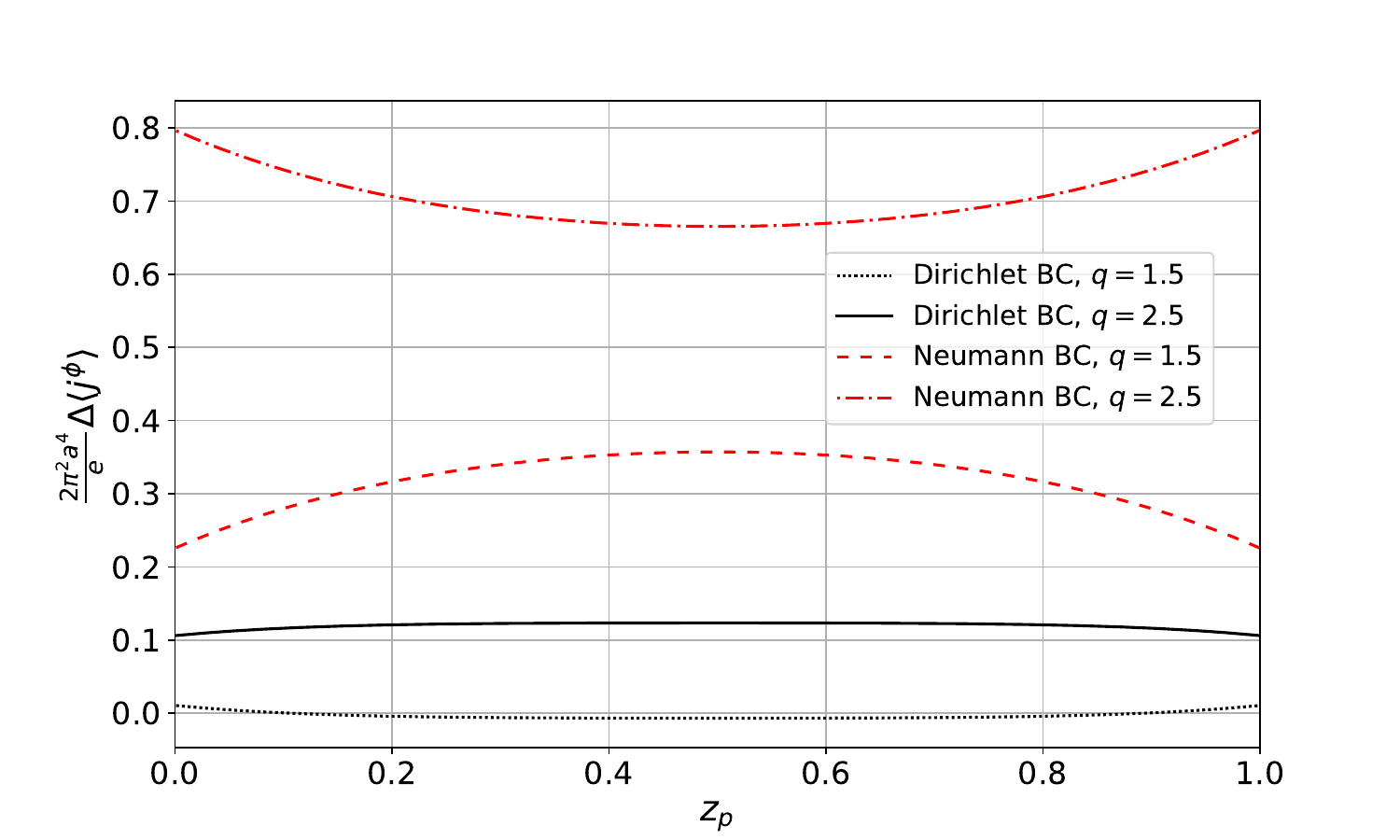}
	\caption{The VEV of the azimuthal current density induced between the plates is plotted as function of $z_p$. In the top panel we assume $r_p=0.1$, and in the bottom, $r_p=0.5$.  For both plots we also assume $D=3$, $\alpha_0=0.25$,  $\xi=0$, $ma=1.5$.}
	\label{fig3}
	\end{center}
\end{figure}

In Fig. \ref{fig3} we display the behavior of the VEV of the azimuthal current density in the region between the plates as function of $z_p$. In the top panel we consider $r_p=0.1$ and in the bottom panel, $r_p=0.5$. Here also, we assume Dirichlet and Neumann boundary conditions and different values for $q$. We can observe from both plots that the VEV is finite on the plates at $z=0$ and $z=1$, being symmetric with respect to the midpoint between the plates at $z=0.5$. Moreover, in both plots the intensities increase with the parameter $q$ and are higher for Dirichlet BC compared with the Neumann BC for the same values of $q$.  

\section{Conclusions}
\label{Conc}
The main objective of this work was to investigate the vacuum bosonic current induced by the presence of a carrying-magnetic-flux cosmic string in a $(D+1)$-de Sitter spacetime considering the presence of two flat boundaries perpendicular to it. In this setup, we impose that the scalar charged quantum field obeys the Robin boundary conditions  on the two flat boundaries. The particular cases of Dirichlet and Neumann boundary conditions are study separately. In order to develop this analysis, we presented the the Wightman function in \eqref{WF7} in a more symmetric for, i.e., decomposed in a part associated with the presence of the string in dS space only, plus the contributions induced by just one flat plane followed by other induced by two flat planes. Because the  current induced by a cosmic string have been calculated before, our focuses were in the obtainment of the azimuthal component induced by a single plate, develop in subsection \ref{Azimuthal}. The contravarient component, $\langle j^\phi\rangle$, was presented in \eqref{CDpl4} combined with \eqref{F-function}, \eqref{u_function} and \eqref{sc}. Some limiting cases for this current have been presented. In the massless conformal coupled was given in \eqref{CDplm0}. For points near the string, and considering $z\neq a_j$, we have shown that for $q|\alpha_0|>1$ this component is finite, and we can take $r=0$ in \eqref{CDpl4}; however for  $q|\alpha_0|<1$ this VEV diverges with ${r_p^{-2(1-q|\alpha_0|)}}$ as shown in \eqref{CDpl-approx}. For points far from the string, \eqref{CDpl4} decays with $r_p^{-(D+2-2\nu)}$. For points close to the plate, $z=a_j$, but outside the string, $r\neq0$,  the VEV is finite; however on the string, $r=0$, and $q|\alpha_0|>1$ the VEV diverges as exhibited in \eqref{CDpl-near-aj}. The Minkowskian limit, i.e., $a\rightarrow\infty$ and fixed value of $t$, has been also considered and is given in \eqref{CDpl-Mink}. Finally for points distant from the plate, $|z-a_j|\gg\eta, \ r$, the current decays as $|z_p-a_j/\eta|^{-(D+2-2\nu)}$. Also in the subsection \ref{Azimuthal}, we have presented two plots, in Fig. \ref{fig1}, exhibiting the behavior of $\langle j^\phi\rangle$ as function of $r_p$ (top panel) and $z_p$ (bottom panel), considering separately the Dirichlet and Neumann BC and different values attributed to $q$.  We can observe that these plots are in accordance with our asymptotic analysis.

The analysis of the VEV of azimuthal current in the region between the plates, has been developed in subsection \ref{CD-two-plates}. The complete expression for this VEV is given in \eqref{CDInt3}, combined with \eqref{v_variable_1} and \eqref{v_variable_2}. Some limiting cases for this contribution has have been analyzed. For a conformal coupled massless scalar field case, the VEV takes a simpler form given by \eqref{Conformal-two-plates}. In the asymptotic limit of large values of the distance between the plates, $\tilde{a}=|a_1-a_2|\gg r,|z-a_j|$, it decays with the inverse of $(\tilde{a}/\eta)^{D+2-2\nu}$as shown in 
\eqref{Limit_a_large}. For large distances from the string and considering fixed distances from the plates, $r\gg \eta,|z-a_j|$, the corresponding asymptotic formula is present in \eqref{Limit-large-r} and shows that the VEV induced in the region between the plates decays as $1/r_p^{D-2+2\nu}$. The Minkowskian limit has been also analyzed for this contribution and it is presented in \eqref{Two-plates-Mink}. The behavior of the VEV of the azimuthal current density in this region, as function of the proper distance from the string, $r_p$, considering Dirichlet and Neumann boundary conditions with different values of $q$, are exhibited in Fig. \ref{fig2}. In the same region, in Fig. \ref{fig3} we have plotted the behavior of the azimuthal current density the proper distance from the plates, $z_p$, considering also the same boundary conditions and different values of $q$. Like in the previous graphs, the plots confirm the analytical asymptotic behaviors.

To finish this paper we want to say that in our analysis we have considered the spacetime fixed. In this sense we have quantized only the matter field. Here, the charged bosonic field. The induced azimuthal current can be considered as the source in the semiclassical formulation of the Maxwell equations. By its turn, the energy density present in the corresponding electromagnetic field can also be considered as source in the Einstein equation in a back-reaction approach, providing corrections on the metric tensor of spacetime background. The calculations of these corrections correspond in fact a hard work that can be developed in new project.

\section*{Acknowledgments}
We want to thank A. A. Saharian for helpful discussions. W.O.S. is supported under grant 2022/2008, Paraíba State Research Foundation (FAPESQ). H.F.S.M. is partially supported by CNPq under Grant No. 308049/2023-3.

\end{document}